# ALGEBRA OF FUNDAMENTAL MEASUREMENTS
## AS A BASIS OF DYNAMICS OF ECONOMIC SYSTEMS

[1]I.G.Tuluzov, [2]S.I.Melnyk

[1] Kharkov, Kharkov Regional Center for Investment
[2] Kharkov, Kharkov National University of Radio Electronics

**Contents**

## Introduction

In the existing economic models, the laws of dynamics are obtained either empirically, based on the analysis of statistic data, or phenomenologically, based on the a priori information about the system. In this case, the proprietor's or the consumer's consciousness cannot be modeled using this method. The known methods of considering its properties (in the theory of consumer function, for instance) are based on the assumption of "rationality" of behavior and these methods are also phenomenological. Models of consciousness based on the analysis of physical mechanisms of mind, which are being actively developed today, are still far from practical application due to complexity and individuality of these mechanisms.

In this paper, we propose an axiomatic approach to constructing the dynamics of systems, in which one the main elements is the consciousness of a subject. At the same time, the main axiom is the statements that the state of consciousness is completely determined by the results of measurements performed on it. This paradigm corresponds to the methodology of constructing physical quantum models, in which the state of object is completely and unambiguously determined by the complete set of the observed variables. However, while in physics a comparison with etalon can be considered a fundamental measurement (N. Bohr), in case of economic systems we propose to consider an offer of transaction as a fundamental measurement. Refusal of consent of the proprietor not only characterizes its state, but also changes it, as it has the form of an economic obligation given to a partner or received from him.

Logical operations "and" and "or" allow constructing more complex transactions based on monatomic (not representable as composite) offers. Standard axioms for such operations allow constructing Boolean algebra based on a set of transactions. However, the possibility of concluding transactions with delayed choice (futures and options) required a reconsideration of this axiomatics. In particular, we have previously obtained an economic analog of Bell inequalities and it has been shown that in the general case for the description of sequence of results of economic measurements the application of quantum-mechanical formalism is required [1]. In papers [2,3] we have discussed the economic analog of the slit experiment. It has been show that the solution of contradictions arising in the process of classical description is possible either by means of expanding the set of transactions (consideration of incomplete transactions), or by introducing the quantum-mechanical description.

Transactions with delayed choice, discussed in this paper, represent a logical generalization of incomplete transactions and allow for a more rigorous approach to studying the properties of the algebra of economic measurements.

Further development of the axiomatic approach requires constructing the geometry of economic measurements and economic states and derivation of equations of dynamics in the form of variational principle in accordance with the similar approach in physics (the theory of selective measurements of J. Schwinger [4]). Considering the behavior of an object as a sequence of actions and choices allows extending the obtained results to random systems, which include the subject's consciousness as one of its elements.

It should be noted, that nowadays a mathematical formalism allowing generalization of the classical theory of probability and logics in such a way that it could include quantum probabilities as well, is being actively developed.

These are the superalgebras of non-commutative operators in [5, 6], for example. Attempts are made to connect the conclusions of these theories with various mechanisms of consciousness (memory, etc.) [7, 8]. In fact, these theories represent a more rigorous approach, compared to Schwinger's approach, with no linking to any physical models and spaces. We are not aimed at the development of this mathematical formalism. However, we will demonstrate that the natural description of the dynamics of economic systems (as well as the subject's consciousness) in the framework of the algebra of fundamental measurements necessarily results in such formalism. In turn, the specifics, on which the description of fundamental measurements is based, allows easily transferring the formalism of Schwinger's theory of selective measurements (and the consequent quantum-mechanical formalism) to the economic and social systems.

## 1 Quantum-mechanical methodology of model description in the theory of consciousness and in economics

### 1.1 Description of system dynamics in econophysical models

Most of the existing econophysical models can be attributed to one of the two classes. The first class uses the <u>empirical approach</u>. In this approach, the dynamic model of the system related to a certain class of integral-differential equations is identified on the basis of the analysis of statistic data (exchange prices, indicators of market state) and the parameters of its description are determined. In fact, this approach uses the "black box" model. Laws of the system's functioning are derived not from its structure, but rather on the basis of the analysis of the previous "trajectory". In this case, physical approach helps using the already known solutions for the classes of equations well-studied in physics previously. The main disadvantage of such approach is the impossibility of selecting the appropriate class of model for sufficiently complex systems, as well as the ambiguity of interpretation of the observed system dynamics.

In contrast with the aforesaid method, the second type of models uses the <u>phenomenological approach</u>. In this approach, a certain mechanism of system's behavior is developed on the basis of a priori information about the system, or even plausible assumptions (properties of the consumer or production function). The parameters of the obtained model are also determined by means of comparison with the statistical data of its observation. Among the disadvantages of this approach is the speculativeness of assumptions, on which such models are based. For instance, most models of agents use a certain speculative mechanism of their interaction, which is actually derived from the phenomenological macrodescription of behavior of each of them and interconnections between them.

Both these approaches have their own specific features and provide solutions for a number of practical tasks. Nevertheless, in both approaches it is practically impossible to model the proprietor's consciousness, which imposes significant restrictions on their range of application. In the present paper we are making a special emphasis on construction of a correct model of consciousness and research of its contribution to various economic models.

### 1.2 Methods of consciousness modeling

• <u>Physical (neurochemical) approach</u>

This approach most fully corresponds to the spirit of exact sciences and is being actively developed today, as a result of breakthrough both in the sphere of nanotechnologies and in quantum informatics. It is usually believed that the higher nervous activity is performed on the level of neural networks, chemical reactions, i.e. rather large macroscopic objects, while the quantum mechanics is something adequate only for the microcosm. However, the discovery of mesoscopic objects with quantum properties allowed assuming a significant role of quantum processes in human consciousness. Nevertheless, models of consciousness based on the analysis of physical mechanisms of thinking (including quantum models – R. Penrose [9] and others) are still far from practical application due to their complexity and individuality of these mechanisms.

•<u>Phenomenological approach</u>

This approach directly deals with macroscopic effects in human consciousness and the model of behavior is developed with no relation to its physical mechanisms. The main disadvantage of such approach is the significant dependency of the obtained models on the initial assumptions on the "rationality" of consciousness. At the same time, the influence of knowledge of the results of such modeling on the consciousness is not taken into account.

•<u>Axiomatic approach</u>

In this approach connected with the methodology of quantum macroscopic games [7] the influence of "questions" on the subject's state is taken into account. It shows that quantum strategies are advantageous compared to classical strategies. Unlike true quantum games (as illustrated in Fig.1), they lack the physical carrier of non-local interaction (quantum bit). Nevertheless, the description of optimum strategies requires application of the quantum-mechanical apparatus.

Fig.1 illustrates the examples of application of each of the three approaches in the process of consciousness modeling, taken from various scientific sources. In the present paper we are going to develop the axiomatic approach. However, unlike the known papers, we will emphasize on the phenomenon of measurement as a basis for constructing dynamic models in exact sciences. In classical physics a measurement is usually considered a secondary procedure, which does not influence in any way the properties of the observed system. However, in the special theory of relativity is it found that the equations of motion can be obtained only with rigorous analysis of the method of measurement of its variables (coordinates and time). Furthermore, in the general theory of relativity, such properties of space and time as curvature, gravity force, etc., are also determined with account of the invariability of measurement results in various reference systems.

However, the influence of measurements on the properties of the observed objects is revealed to the full extent in quantum mechanics. We assume that both in models including a proprietor, and in general models of consciousness, the influence of measurement is determinative. Therefore, construction of such models requires the application of the consistent approach used in modern physics.

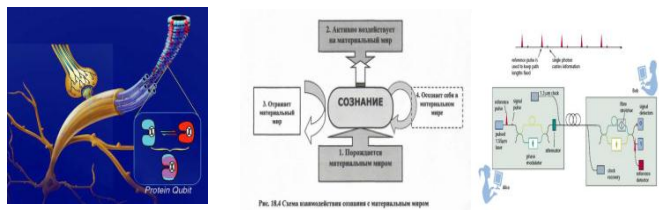

Fig.1. Examples of various approaches to the modeling of consciousness: a) physical, b) phenomenological, c) axiomatic

### 1.3 Grounds for using quantum models of consciousness

Before proceeding to the axiomatics of measurements in the theory of consciousness, let us adduce several arguments stressing the necessity of such approach.

• <u>Paradox of self-description (butterfly effect)</u>.

This paradox is referred in various interpretations in many sources (for instance, Russell paradox). However, it is best illustrated in the attempts of predicting future. The classical model of the world (and consciousness in it) can be arbitrarily accurate as long as there is no prediction made on the basis of this model. After the subject reveals his possible future, his consciousness changes so greatly, that the prediction itself becomes incorrect (Fig.2). However, in this case, taking into account the possible incorrectness of the prediction, the subject's state may change not so significantly (depending on the degree of his confidence in the predicted future). From this argumentation it follows that only a fuzzy description of the subject's state by the subject itself is possible, while an accurate prediction of his own future by himself is impossible in principle.

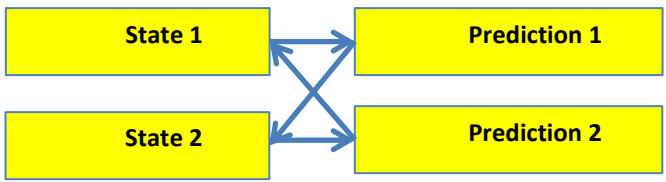

Fig.2. In the attempt of self-description by the subject of his own state for predicting the future behavior, a steady logical link of argumentation can appear, allowing two or more alternatives of behavior.

Let us emphasize the principal difference of physical models from the models discussed in this paper. In physics, the observer is always present in any dimension, but he defines all his actions in advance and does not interfere in the course of experiment later. The main requirement to measurements is their complete reproducibility and objectiveness. At the same time, the essence of measurements in economy and in the theory of consciousness is the choice of observer (subject). Its state is the subject of observation. Therefore, from the one hand, describing the measurement "from inside", as seen by the observer, we have to introduce the category of choice, which is absent in the physical theory. On the other hand, an external observer sees only the statistics of behavior of subjects. Therefore, his description of their behavior is of statistical nature. Below we are going to prove that such "external" behavior can be based on the quantum-mechanical formalism.

• Analog of the principle of uncertainty.

In the simplest case the price of futures contract for buying a certain share (right to buy this share in the future at the current price) has the meaning of the speed of price changing of this share. However, the knowledge of the exact price of such futures contract means that we can precisely predict the share price at the moment of its redemption. In this case, such riskless share will cost a predicted price at the current moment. A contradictive situation occurs which results in "freezing" of the share's price in case of its constant and accurate measurement. In quantum mechanics a similar phenomenon is observed experimentally (quantum Zeno effect). For an adequate description of the dynamics of exchange prices we need to assume that the simultaneous accurate measurement of the share price and the futures contract for buying this share is impossible in principle.

• Economic analog of Bell inequality.

In a model allowing inaccurate values of price of a certain active the possibilities of selling (or buying) it at two different prices are interconnected. The closer the prices are, the closer the possibilities are. This situation qualitatively resembles the EPR experiment, in which the possibilities of obtaining similar results of measurement of spin projection become closer as the directions of measurement of these projections become closer. The general properties of this probabilistic dependency can be obtained from the considerations of symmetry and consistency. In physics they are formulated in the form of Bell inequalities, from which it follows that the classical model of hidden parameters contradicts to the experiment, at least for certain sets of three selected directions of measurement of spin projection. We have previously obtained similar conditions (formulated in the form of inequalities), from which it follows that the classical model of hidden parameters contradicts to the results of measurements of probabilities of sale and purchase of a certain active for any three values of price [10].

• Advantages of strategies of the delayed choice.

In physical theory the category of choice is absent in equations of dynamics of the observed systems. It is due to the requirements of objectiveness and reproducibility of measurement results. At the same time, it is practically impossible to exclude it completely from the description of objects, as the measurement itself changes the state of the observed system. At the same time, as long as the choice of this method is not determined by the observer, we have to describe the dynamics of state as a set (superposition) of alternatives.

In economic models and in the theory of consciousness the subject itself is a part of the observed system. His choice is included in the description of the dynamics as a parameter. Situations are possible when his choice is delayed up to the moment when the absence of choice will contradict to the rest of the measurement results.

Thus, for instance, in the well-known game "Sea battle" a cheating player can move the already deployed "ship" into a free slot of the playing area after a successful "shot" of the opponent (Fig.3). At the same time, very quickly he realizes that it is even more "advantageous" not to deploy the ships at the beginning of the game and to delay his choice until only one alternative of their location remains. In this case, they are placed in the most advantageous position with account of the moves that his opponent has already made.

The description of the player's state in this case is possible not by means of the classical distribution of probabilities of the ship locations, but by means of the set of equally possible alternatives (from the point of view of the subsequent choice). The principal fact is that the best opponent's strategy should be based on the alternative model of state, rather than the probabilistic model. In this case, it is aimed not at hitting a "ship", but on maximally quick reduction of the set of possible alternatives of the opponent to a single possible alternative. The phenomenon of coherent state of consciousness on the example of this game and other argumentation is discussed in detail in our paper [11].

A similar situation in physics is rather well studied and is connected with the discussion of the Wheeler's "delayed choice" experiment.

Thus, the aforesaid examples allow assuming that the description of the dynamics of a system, in which a subject with a possibility of choice is present, requires using the quantum-mechanical formalism. The provision is being used more and more frequently in econophysical models.

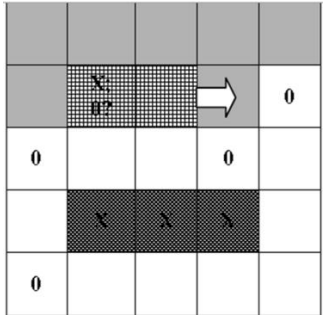

Fig.3. Illustration of the advantageous strategies of the "delayed choice" and forming of alternatives in the player's consciousness on the example of the game "Sea battle"

However, in case of formal using of physical models the economic specifics of the discussed systems is often lost. That is why our approach is based on using only the physical methodology instead of using readily available models. We have previously used this approach for analyzing the possibility of using the physical methodology for constructing classical models of economic systems [12] (Majorana prize 2010).

### 1.4 Mathematical description of the "delayed choice" situation in physics, economics and in the theory of consciousness

So far we have being discussing the choice only in relation to the subject's consciousness in economic models. Nevertheless, it can be formally used in physics for discussing quantum models as well. This will allow us clarifying the meaning of quantum categories in the theory of consciousness and in economics, and proposing an alternative language for the description of dynamics of physical systems.

Let us note that some authors propose considering the phenomenon of choice in human consciousness as a collapse of the state of superposition of several alternatives towards one of them. For the description of this collapse a number of mechanisms based on the nanostructures of fine organization of neurons are used [9]. On the contrary, in the present paper we consider the subject's choice of one of the alternatives the primary element of description and the quantum dynamics of the observed system is obtained as a result of the symmetry properties of this phenomenon. Below we are going to draw parallel between the meaning of various variants of choice in the theory of consciousness and various variants of measurements in quantum mechanics.

CONSCIOUSNESS
- Choice is made and it is known (pure state of consciousness) – determinism of behavior (actions) – classical logics
- Choice is made but it is unknown (mix of pure states of consciousness for a set of subjects) – mixed strategy of behavior – classical theory of games
- Delayed choice (superposition of alternative pure states) – conditional strategies of behavior - quantum macroscopic games
- Dynamics of the system – sequence of choices and actions.

PHYSICS
- Measurement is made and it is known (pure classical state) – classical determinism – one predetermined trajectory in the phase space of states.
- Measurement is made but its result is unknown (mix of pure classical states) – functions of distribution of probabilities in the set of possible alternative trajectories – Liouville equation. Weight coefficient of each alternative is retained.
- Measurement is partially performed (allows various alternatives in the process of subsequent refinement of the result) – the state is described by the superposition of alternatives – the Schrödinger equation for wave function.
- Mathematical description – retaining of complex weight coefficients of each alternative in case of evolution without measurements. The projective postulate determines the probability of a specific result (alternative) in the process of measurement.
- Dynamics of the system – sequence of instant measurements and evolution of state without measurements between them.

Further we are going to illustrate all our argumentations on the example of economic models. According to their specific features, they occupy an intermediate place between physical models and general models of the theory of consciousness. On the one hand, the offer of transaction represents a situation in which the subject (proprietor) has to make a choice. On the other hand, it is nothing else but measurement of his state. The subject's documented answer for an offer is an additional financial obligation, which obviously changes both the subject's state and his further behavior.

Thus, we are going to consider the offer of transaction of property exchange as an equivalent of physical measurement in economic models. And the consent or refusal of the transaction with corresponding financial obligations will be considered as the result of the elementary measurement.

### 2 Algebra of transactions as a mathematical structure for the construction of dynamics of economic systems

### 2.1 Principle of measurability as applied to the theory of consciousness and economics

In the methodology of constructing physical quantum models the state of object is completely and unambiguously determined by the set of the observed variables (results of their measurements). We assume that the state of subject (in economic models – the state of proprietor) is completely determined by the results of measurements performed on him. In this connection the situation of decision-making in economy is more specific, as each such decision is usually documented. In the general case of human behavior the adopted decision can be of various degrees of certainty of its irreversibility. The state connected with unclearly made decisions can be described similarly to the theory of continuous fuzzy quantum measurements, which is being actively developed today in connection with measurements of quantum bit states.

In physics the fundamental (elementary) measurement is the comparison of object's state with the standard (N. Bohr). We propose to consider the offer of transaction as such fundamental measurement in economic systems. The results of the offer of transaction are the financial obligations or rights depending on refusal or consent for the transaction. We will also assume that the economic state of proprietor is completely determined by the availability of obligations and rights obtained both in the process of receiving an offer of transaction and in the process of giving an answer to this offer.

This statement is principal for the whole further modeling, we will refer to it as the principle of measurability of state.

We realize that in any real situation two subjects (even with equal financial obligations) can choose different strategies of behavior. However, such situation is encountered in physics as well. There is no need to ensure full identity of objects for the reproducibility of the performed experiments. It is only sufficient to ensure the identity of conditions and properties essential for this specific task. In case if the set of identical properties does not ensure the identity of behavior of subjects, it can only mean that we have not taken account of all their properties significant for the discussed problem.

Thus, the principle of measurability and the idealization of optimum behavior of proprietors ensure the equivalence of their states (and the related behavior) in case of availability of equal packages of obligations and rights at their disposal.

## 2.2 Economic idealizations forming the basis of the axiomatics of the algebra of transactions

Following the aforesaid logic of theory constructing, we are not going to use the readily available mathematical structures well known from physics. For their construction we are going to start with several economic assumptions of axiomatic nature. These assumptions are going to determine all the properties of the eventually obtained models. Nevertheless, we will strictly follow the sequence of constructing equations of dynamics, generally accepted in exact sciences in general and in physics in particular.

• **The property of completeness of choice** means that a refusal of an offered transaction automatically becomes an obligation (consent) for making a reverse transaction. If, for instance, an exchange broker refuses to buy a share for 100 Euro, it means that he agrees to sell it at the same price, and vice versa.

• **The property of absorption** means that if a proprietor agrees to sell his property for $n$ conventional units, then he agrees to sells it for any larger quantity of these conventional units. If he agrees to buy something for $n$ conventional units, then he agrees to buy it for a smaller quantity of units as well.

• **The property of local equity** (absence of arbitration in the proprietor's state). This property refers to the consistency of simultaneous preferences of the proprietor. If, for instance, he agrees to exchange value A for value B and value B for value C, then he refuses to exchange C for A.

These economic assumptions form the informal essence of the proposed axiomatics of fundamental measurements in economics.

In accordance with these properties we will consider the fundamental measurements of the state of proprietor as elementary offers of transactions of property exchange. In relation to buyer-seller, the transactions of sale and purchase have the symmetry described below:

1. Proprietor «A» offers to proprietor «B» to buy a certain valuable item X for a1 conventional units of another valuable item Y.
   • This means that «A» provides obligations of exchanging X for a1*Y in case of consent of B.
   • This also means that «B» receives the right to exchange a1Y for X in case of his consent.

2. Proprietor «A» offers proprietor B to sell a certain valuable item X for a1 conventional units of another valuable item Y.
   • This means that «A» provides obligations of exchanging a1*Y for X in case of consent of B.
   • This also means that «B» receives the right to exchange X for a1Y in case of his consent.

Let us note that any of the elementary offers of transaction is at the same time an obligation for one of the proprietors and a right for the other proprietor (two copies of agreement are concluded). Therefore, in an enclosed system this property derives the law of conservation of the summary right of choice for delayed transactions (transactions which have already been offered, but no decision on refusal or consent for the transaction is made yet).

## 2.3 Algebra of elementary offers of transactions

Let us introduce a binary operation of addition of two transactions. We will refer as $M_1+M_2$ to the transaction, in which the package of documents includes the packages of transactions $M_1$ and $M_2$ and an additional condition: the proprietor refuses from the transaction $M_1+M_2$ if he refuses from at least one of them, and accepts it if he accepts both transactions, having preliminarily become familiar with the content of both transaction. Hereinafter we will refer to the refusal of the transaction as "1" and the consent for it as "0". This is done in order to provide the truth tables for operations «+» and «*» corresponding to the generally accepted logical operations "and", "or". At the same time, there can be a certain confusion as in our interpretation "1" refers to a refusal rather than a consent. The meaning of such designations we become clear later, in the process of discussing screen illustrations to the algebra of elementary offers of transaction.

Let us introduce a binary operation of multiplication of two transactions. We will refer as $M_1 * M_2$ to the transaction, in which the package of documents includes the packages of transactions $M_1$ and $M_2$ and an additional condition: the proprietor refuses from the transaction $M_1 + M_2$ if he refuses from both transactions, and accepts it if he accepts at least one of them, having preliminarily become familiar with the content of both transaction.

It can be easily checked that the properties of commutativity, distributivity and associativity are valid for operations determined in this way. Actually they are similar to the operations of addition and multiplication of statements in classical logics.

Let us introduce a unitary operation of negation of transaction. We will refer to transaction $M_2$ as the negation of transaction $M_1$ and denote it $\overline{M_1}$ if the proprietor refuses from the transaction if and only if he accepts the transaction $M_1$.

The examples of such reverse transactions are, for instance, transactions of sale and purchase of a certain valuable item for the same price, if both these transactions are offered to the proprietor simultaneously. This statement represents an idealization of our model. This implies, in particular, that the proprietor will never simultaneously refuse from mutually inverse and will never simultaneously accept them. This property can be written as

$$\overline{M_1} + M_1 = 1; \qquad \overline{M_1} * M_1 = 0 \qquad (1)$$

In general algebra a lattice can be determined as a universal algebra with two binary operations «+» and «*» satisfying the following identities

$M1 + M1 = M1$ (idempotency of addition)
$M1 * M1 = M1$ (idempotency of multiplication)
$M1 + M2 = M2 + M1$ (commutativity of addition)
$M1 * M2 = M2 * M1$ (commutativity of multiplication)
$(M1 + M2) + M3 = M1 + (M2 + M3)$
(associativity of addition)
$(M1 * M2) * M3 = M1 * (M2 * M3)$
(associativity of multiplication)
$M1 * (M1 + M2) = M1; M1 + (M1 * M2) = M1$
(absorption) $\qquad (2)$

The last property of absorption is valid for homogeneous transactions (connected with the offer of exchange for different quantities of the same valuable item). Each of such subsets of homogeneous elementary offers of transactions represents a commutative lattice with 0 and 1.

## 2.4 Origination of quantum properties in the algebra of elementary offers of transactions

The axiomatics of the constructed algebraic structure corresponds to the classical logic of Aristotle, which also represents a commutative lattice. Nevertheless, there is principal difference between them. In the classical logic it is additionally assumed that the validity of each statement is initially determined and does not depend on whether we know about it or not. In the algebra of elementary offers of transactions the specific values of validity of monatomic statements are not required if the validity of the compound statement is known.

In our axiomatics the values "0" and "1" appear only as a result of measurement of the proprietor's state in case of his consent or refusal of the transaction, respectively. This can result in a situation when an answer for the offer of compound transaction (for instance, $M_1+M_2$) is received, but the choice of one of the possible variants of the answer for the offers $M_1$ and $M_2$ separately is delayed. The proprietor's state (supported by the set of his obligations and rights) remains in the state of superposition of possible alternatives before a corresponding (specializing) offer is made.

Let us note that in economic systems the strategy of delayed choice is advantageous as it always allows a possibility of non-delayed "internal" choice without formalizing any financial obligations. Therefore, the optimum strategy of a proprietor is the strategy of retaining to the maximum degree of the alternatives of choice until they start to contradict with the undertaken obligations.

This specific idealization of the optimum behavior of proprietors allows us using the previously formulated principle of measurability of state.

## 2.5 Screen representation of elements of the set of elementary offers of transactions and the generalized indivisible transaction

It is appropriate to discuss the algebra of elementary offers of transactions on the basis of analogy with slit experiments in wave and quantum mechanics. As each offer of elementary transaction presupposes a choice of only two variants of answer – a consent or a refusal, it can be considered an analog of the result of measurement in the experiment of a particle flying over or under a non-transparent semi-screen (Fig.4)

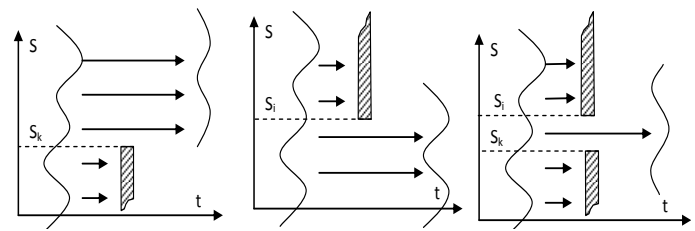

Fig.4. Screen representation of a set of elementary offers of transactions. Slit experiments as measurements of proprietor's state.

a) Lower semi-screen. Offer of purchase of property for the price «Sk» «passes through» those proprietors who refused from the transaction.

b) Upper semi-screen. Offer of sale of property for the price «Si» «passes through» those proprietors who refused from the transaction.

c) Slit in the screen. If a proprietor assumes that his property cost more than «Sk», but less than «Si», the equivalent "particle" passes through the slit.

At the same time, refusal of the transaction corresponds to the situation when a "particle" misses the screen. In the classical model it means that the result of measurement is obtained, but the state of the particle has not changed. In case if the particle possesses quantum properties, then its state changes the stronger the closer to the edge of the screen it passes.

In the economic models the value of the vertical (measured) coordinate of particle corresponds to the proprietor's opinion on the price of his property. In case if his opinion is precise, the refusal of purchase or sale will not influence the state of his consciousness in any way. However, if the initial state is fuzzy (represents a set of alternatives), the refusal of the offered transaction and the obligations undertaken as a result of such refusal change the initial state. The algebra of elementary offers of transactions is mainly aimed at obtaining mathematical formulas for the description of the influence of refusal of the offered transaction on the proprietor's state.

The screen representation of the algebra of elementary offers of transactions, the meaning of hidden parameters of consciousness and other aspects of slit experiments have been discussed in detail in our papers [2,3]. In the general case, the offer of transaction can have a complex structure and consist of elementary offers. Nevertheless, such transaction remained indivisible if only one answer for the offer is required - either a consent or a refusal of the offer of exchange.

With account of the aforesaid properties of algebraic operation, any invisible transaction (regardless of its complexity) can be considered as a formula of Boolean algebra and can be represented as a disjunctive normal form (DNF). In case of combining homogeneous elementary offers of transaction, each conjunction in the DNF can be represented by only two cofactors (slit in the screen) due to the property of absorption. With account of the property of absorption the superposition of two slits transforms into a sum of non-overlapping slits in accordance with the corresponding formula. The final form of any indivisible transaction is equivalent to the final set of non-overlapping slits in an infinite screen (Fig.5).

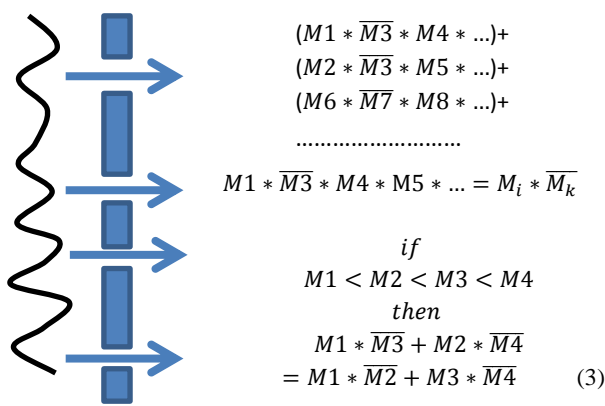

Fig.5. Representation of a generalized indivisible transaction using a set of non-overlapping slits in an infinite non-transparent screen.

The sensor absorbing the particle is modeled as $M1 * \overline{M2}$ if $M1 > M2$, on the contrary to $M1 < M2$, as in case of the slit.

### 2.6 Formation of the set of selective measurements based on the algebra of elementary offers of transactions

The final aim of our study is the derivation of equations of dynamics written in the form of differential or operator equations. In the general case, it requires a number of sequential stages. They include studying the properties of symmetry of the set and construction of the geometry of measurements – the vector space of states, in which the "motion" is going to take place. Then it is necessary to formulate the variational principle, which will be the main "decision rule" determining the laws of dynamics. The limiting process to the continuous description of evolution of state allows deriving the differential equations of motion following from the variational principle.

There is a historically established situation in physics when a number of successive stages of such derivation can be omitted. For instance, the space of states of most of the observed physical systems had been described long before the theory of measurements in this space was formulated. In our opinion, this situation is the main obstacle for the consistent development of physical approach in the adjacent spheres.

Nevertheless, such sequential approach to construction of physical theory (from the notion of generalized measurement to the equations of quantum electrodynamics) was performed in the scientific works of J. Schwinger [4]. Therefore, below we are going to omit a number of calculations common for physical and economic measurements, referring to this paper. The fundamental principles of the theory of selective measurements of J. Schwinger are the following:

• The selective measurement (SM) of a physical value $A$ is a measurement $M(a')$, which passes systems with value $a'$ for property $A$ and discards the rest.

• The operation of addition of selective measurements is a measurement, which selects an ensemble of objects corresponding to all the values of $A$ in the sum, but does not distinguish them. The addition of selective measurements is commutative and associative.

• The operation of multiplication of selective measurements denotes their sequential execution. The multiplication of selective measurements is associative.

• Physical values are called compatible if the measurement of one of them does not violate the information obtained as a result of the preceding measurement of the other one.

• The set of compatible physical values is called complete if each pair of the set is compatible, but any other value is not compatible with each of the elements of the set.

Let us illustrate that a set of generalized indivisible transaction can be considered as selective measurements. The symbol of elementary measurement $M(a')$ corresponds to the offer of a generalized indivisible transaction, in which the proprietor is offered to sell a certain value for $(a_i - \delta/2)$ or buy it for $(a_i + \delta/2)$ conventional units $A$. The proprietor who refuses both offers is "passed through" the equivalent slit with width $\delta$ in the screen, corresponding to the property $A$. All the rest of the proprietors are absorbed by the screen and are not included in the further consideration.

The sequential execution of selective economic measurements is described by the binary operation of non-commutative multiplication. In order to avoid confusion with the operation of commutative multiplication already introduced for the set of elementary offers of transaction, we will denote it with the symbol «·», or simply sequentially write down the symbols of measurement assuming their non-commutative multiplication. Let us note that the algebra of elementary offers of transactions with its algebraic operations is more a fundamental structure revealing the essence of measurements of consciousness state. The algebra of selective economic measurements is based on the set of generalized indivisible transactions rather than elementary ones. Therefore the essence of quantum-mechanical properties of consciousness can only be revealed by means of studying these two structures in aggregate as it is done by us is our paper [3].

### 2.7 Compatible properties. Determination of the economic state

The economic state of the system is primarily determined by its consumer properties. And these properties appear (and can be measured) as a result of various elementary economic events – transactions. Therefore the compatible consumer properties are those properties, which are not interconnected in any way at all. For instance, in case of using diamonds for strengthening the surface of cutting tools their transparency does not influence their price in any way. At the same time in case of using them for the production of optical elements their hardness is practically inessential. But such a quality (consumer value) of diamonds as the crystal size effects the price per carat both in the first and in the second case. Let us consider (very conventionally) that the diamond crystal size determines their consumer value in case of using them in jewelry. Then we can consider that the measurements of value of diamond carat in case of their "instrumental" and "optical" use are compatible, and the measurement in case of their "jewelry" use is not compatible either with the first or with the second measurement. Therefore, by finding out the price per diamond carat in instrumental industry (or receiving an offer of transaction) a customer can change his opinion on their "jewelry" value, but will not receive any information on their "optical" value. Let us emphasize that the term value refers, as previously, not so much to the physical properties of the diamonds, as to their consumer value, measured as a possibility of exchanging them for a certain product in a certain transaction. The received information cannot change their physical properties, but can change the state of consciousness of their owner and his estimate of their cost.

For determining the complete set of variables of the economic state, let us first note that we consider the offers of transaction of exchange for similar values varying only in the quantity of $a_i$ as a different generalized selective measurement of the same variable $A$.

If an exchange for a certain quantity $b_k$ of a qualitatively different value (for instance, dates instead of bananas), then this generalized selective measurement corresponds to a different variable of state $B$.

The degree of dependency of the results of measurements of one of the variables on the known value of the other variable (and the financial obligations undertaken in this connection) depends only on the degree of appreciation by the proprietor of the values $a_i$ and $b_k$ offered for exchange. This degree of appreciation, in turn, can only be determined as a result of additional measurements of inter-exchange of these values for each other. In this connection, we have reasons to assume that the quantity of independent consumer values is limited, and it determines the dimensions of the space of states.

In physical models the equivalent of the compatible properties of quantum particles are the orthogonal vectors in a certain domain of state. In our next publication we will analyze the issue of the geometry of economic states in detail. After determining the complete set of compatible economic parameters it is natural to introduce the notion of **complete economic measurement**. In our interpretation it is a set of transactions, the results of which determine the economic state of the system to the maximum complete degree. It means that no other transaction exists, the results of which do not depend on this state.

Let us also note that, by introducing a measure for a set of generalized indivisible transactions, we can interpret the above-formulated property of "local equity" as an inequality of triangle, and the principle of measurability can be interpreted as a statement that the result of measurement (transaction), not included in the complete set, cannot be predicted more precisely rather than on the basis of the results of measurement of the complete set of variables.

At this point, we consider appropriate to quote a number of considerations of common nature, clarifying in a similar manner the essence of compatible and incompatible measurements both in economics and in physics. It follows from their determination that the result of any measurement, which is not included in the complete set, depends on the system state, which is determined by the results of the complete set of measurements. However, we can state that an even stronger statement is valid: **the result of measurement not included in the complete set is completely determined by the results of the complete set of measurements.** The term completeness refers to the absence of any regularities not following from the results of the complete set of measurements.

Let us suppose that this statement is wrong. Then we can construct a complex measurement consisting of multiple iteration of the same testing, calculation of the statistical parameter characterizing the new regularity (not connected in any way to other results of the complete set of measurements) and forgetfulness (circulation) of the remaining information received in the process of testing. Such a composite measurement will be compatible with the rest of the measurements of the complete set. By adding it to the complete set we can ensure the validity of the stronger statement for all the remaining measurements. A more detailed substantiation of this thesis is represented in paper [11].

### 2.8 Economic meaning of the transformation function

#### 2.8.1 *Economic measurements changing the system states*

Both in physics and in economics, it proves insufficient to use only the selective measurements of slit or semi-screen type for the description of the system dynamics. In the process of such economic measurements the only subject of changing is either the proprietary right for a certain product or the proprietor's opinion on its value. At the same time, the actual consumer properties change only as a result of utilization of these products. For commonality, we will refer to all such processes as the **technologies,** or measurements changing the state of the observed system (following the Schrödinger's terminology). Our aim is to show that these two types of measurements are sufficient for the description of the dynamic events in economic systems by the analogy with physics.

Let us note that in the "state-changing measurement" no selection of incoming elementary particles occurs. They all have different input and output states. Thus, the division of measurements into selective and state-changing allows representing all the observed dynamics as a structure consisting of such measurements. A question arises in this connection – to what extent this approximation is valid. In physics the approach based on the theory of continuous measurements is being increasingly used recently. In these measurements the processes of information reception, selection and changing of states occur continuously in time and parallel to each other. In our opinion, economic systems correspond to discrete description to a greater degree compared to physical systems. It is due to the fact that the exchange procedure is always documented by the transaction, while a continuously concluded transaction is nonsense. Nevertheless, in case if such measurements are performed sufficiently often, we can also replace their aggregate set by a single continuous measurement. This approximation has allowed us obtaining the generalization of the Black-Scholes formula for this specific case [13].

Measurements changing the system state are an integral part of any economic process. The exchange procedures (selective measurements) lose their sense if the resources received as a result of such exchanges will not be used for changing of state of their proprietor. In this context even an eaten hamburger is a technology changing the state of an elementary economic object (subject who ate this hamburger and who was adopting certain economic decisions).

#### 2.8.2 *Transformation functions of economic measurements*

Following the logic of statements in paper [10] we can consider the sequence of the technology $M(b',c')$ and the selective measurement $M(a')$ as a certain technology $M(a',c') \approx M(a')M(b',c') \equiv M(a',a')M(b',c')$, which selects the incoming particles in state $c'$ and transforms them into state $a'$. At the same time at the stage of selective measurement (selection) of particles possessing the property $b'$ only part of them pass through the "economic price slit" $a'$. Therefore we can conditionally write down the following

$$M(a')M(b',c') = \langle a'|b'\rangle M(b',c'), \quad (4)$$

where $\langle a'|b'\rangle$ is a certain factor with yet undetermined meaning. These are the proprietors who, for instance, evaluate their property for $b'$ units of grain and at the same time accept a payment in the amount of $a'$ units of meat for it. Let us compare these proprietors with those who, for instance, considered the price of their property equal to $d'$ units of vine after the first selective measurement. The compound measurement responsible for their selection can be written down as

$$M(d',c')M(b') = \langle d'|b'\rangle M(a',c') \quad (5)$$

At the same time, it is possible that we can have the same number of proprietors in the same state $c'$ at the input of measurements (4) and (5) and the same number of proprietors, who had agreed that the cost of their working day is $a'$ at the output. But will the properties of these proprietors (relative to the subsequent measurements) be the same?

By making formal calculations, we will take into account that $\sum_{a'} M(a') = 1$. $\sum_{b'} M(b') = 1$. These equalities only mean that the subject will surely agree for the transaction for a certain quantity of product *a* or product *b*. Then we can write down the following

$$M(c',d') = \sum_{a',b'}\langle a'|c'\rangle\langle d'|b'\rangle M(a',b') \quad (6)$$

The economic meaning of the equality is that any technology $M(c',d')$ can be represented as the sum (by not as a mixture) of technologies $M(a',b')$ by going over various pairs $(a',b')$. By summarizing in (6) all the possible values of $(a',b')$ we are actually operating with packages of obligations corresponding to each of them. The sum of symbols of measurements, as above, only means the formation of a new package of offers from the summands according to specific rules. The term "sum of technologies" cannot be interpreted as a mixture of products of these technologies or as their simultaneous application to the same raw material (such explanation can prove to be completely impossible). The equality $M_1 = M_2 + M_3$ only denotes the rule, according to which we can calculate how the consumer properties of the object are changed as a result of measurement $M_1$, knowing how they are changed as a result of measurements $M_2$ and $M_3$. At the same time, let us note once again that the term "consumer properties" refers only to the measurable parameters in the fundamental economic measurements (probability of consent or refusal of a certain transaction). The first part of the inequality (6) is no more than a different form of notation of its left part. However, <u>assuming</u> that the symbols a'c' and others are cofactors, the **fundamental property of composition of transformation functions** can be obtained for them:

$$\sum_{b'}\langle a'|b'\rangle\langle b'|c'\rangle = \langle a'|c'\rangle \quad (7)$$

Summarizing of various values of $b'$ formally means that in the interval between the transactions (selection of proprietors) in states $c'$ and $a'$ we are offering them to make a deal for the sale of their property at any of the possible prices $b'$, which they are not even obliged to announce. It is obvious that all proprietors will agree with this, and this consent will not change state and will not add any information about them.

Thus, the fundamental property of the transformations is trivial if the summation is made with all the possible values of the complete set b, however it allows explaining the paradox quantum properties when the set of acceptable alternatives is limited (for instance, in a two slit experiment).

The mathematical essence of this fundamental property is that if a', b' or c' are the values of complete sets of compatible variables, then the laws of transformation of these values from a' into b' and from b' into c' allow calculating the corresponding law of transformation from a' into c'. It also appears that the representation of transformation symbols by complex numbers is sufficient to:
- Satisfy the fundamental property of composition
- Ensure the validity of properties of symmetry for the discussed subset of the generalized selective measurements
- Connect these numbers with the experimentally measurable probabilities of penetration of particles (or probabilities of refusal of packages of offers in economic models).

Thus, for instance, in Stern - Gerlach experiments various complete sets of variables correspond to various directions in physical space. It can be easily shown that the only function satisfying the aforesaid properties is the function $\cos(\alpha/2)$, where α is the angle between these directions. If the properties of symmetry of a set of dimensions similar to this experiment will be detected in a certain economic system, then the corresponding states can be connected with the set of directions in the virtual economic space of states.

Let us note the formal analogy of the fundamental property of composition with the laws of transformation of various reference frames in classical mechanics. The results of the complete set of measurements represent the projections of the economic state of the set of objects on the corresponding reference system, and the set of symbols $\langle a'|b'\rangle$ represents the coupling coefficients of the descriptions of the same economic state in different reference systems.

It is worth making a remark on the **relative completeness** of the set of consumer properties described as $a'$, $b'$ or $c'$. It is determined only for the subset of consumer properties appearing in the discussed economic events (employment of workers). In case of taking account of a wider spectrum of properties connected with other transactions, these sets can turn out to be incomplete. At this point, the analogy with the measurements of electron spin projection in the experiments using the Stern-Gerlach device is appropriate. In case of measuring this projection in the plane perpendicular to the direction of movement of the electron, the complete set of selective measurements can be formed by any two perpendicular directions in this plane. At the same time, this set is incomplete relative to the measurements of the spin projection in all the directions of a three-dimensional space.

### 2.8.3 **Statistical meaning of the transformation functions**

We have previously stated that the result of fundamental measurement (comparison with the etalon) is the positive or negative answer. In economic application it is represented by consent or refusal of the offered transaction. The results of the complete set of such offers unambiguously determine the state of proprietor. Therefore these and only these results can determine the results of subsequent measurements of variables of a different complete set. However, in the general case the interconnection of these results is not deterministic. Therefore, the next natural generalization of the fundamental measurements is the measurement of the probability of obtaining a certain result for a set of objects being in a specific state. The aforesaid coefficients of the function of transformation from one complete set of measurements to the other are to determine these probabilities.

The general paradigm of the theory of fundamental measurements is the statement that the final result of measurement must be a number. In case if the result of measurement has a dimension (bananas or dates, as we have illustrated earlier), then its qualitative properties require an additional description. However, for obtaining a non-dimensional number we must compare the quantities of objects in equivalent (indistinguishable) states. Therefore, the initial fundamental results of measurements for each pair of values of a complete set of variables are the transmission coefficients appearing in the process of generalized selective measurements of the following type

$$M(a')M(b')M(a') = \langle a'|b'\rangle\langle b'|a'\rangle M(a'),$$

or $M(b')M(a')M(b') = \langle b'|a'\rangle\langle a'|b'\rangle M(b')$ (8)

In the first case the number $\langle a'|b'\rangle\langle b'|a'\rangle$ is the experimentally measurable probability of detecting the object in state a' of the complete set of variables a, if it is in the state b'. In the second case, the situation is inverse.

We can also consider an infinite sequence of measurements

$$M(b')M(a')M(b')M(a')M(b')M(a')\ldots =$$
$$\langle a'|b'\rangle\langle b'|a'\rangle\langle a'|b'\rangle\langle b'|a'\rangle\langle a'|b'\rangle\langle b'|a'\rangle\ldots M(b'),$$

By grouping the coefficients of the transformation function in various ways, it can be shown that

$$\langle a'|b'\rangle\langle b'|a'\rangle = \langle b'|a'\rangle\langle a'|b'\rangle = \rho(a';b') \quad (9)$$

There are other physical and mathematical arguments for the probabilistic interpretation of these coefficients. In particular, in physics the interconnection between the symbols of the function of transformation and the statistical regularities can be based on the properties of invariance of these symbols in relation to their multiplication by a random phase factor.

In the economic models the notions connected with probability and sets of alternatives more transparent. It has allowed us to show [3] that the quantum laws of addition of probabilities for alternatives represent a simplified description of a more complete set of experimental results, in which the parameters of state of both the observed economic particle and the observer are taken into account.

### 2.9 Proceeding from the measurements in economic models to the theory of consciousness

In this paper, we have been so far discussing only the economic models and the possibility of application of the theory of selective measurements to them. However, the specifics of economic relations has been formalized by us in a rather general sense, as a change of consumer properties of economic elementary particles both as a result of transaction (or refusal of it) and in the process of a certain technological changing.

It is easy to note that the human behavior in the general case can also be described using these terms. We can consider a subject's consciousness as an elementary particle of consciousness. Then the selective measurement of state of such particle is a result of selection of a specific type of behavior under the specific circumstances. The set of these circumstances forms the terms and conditions of the transaction (with the corresponding "payments" depending on the adopted decision). If the subject's choice is limited to only one of the two alternatives – consent for a certain action (1) or refusal of it (0), then such measurement can be considered elementary.

By analogy with economics, among the infinite set of possible actions both compatible (decision on one of them is not connected in any way with the decision on the other action), and incompatible actions exist. The latter are immeasurably more numerous, but the circumstances determining them (terms and conditions of the transaction) can be expressed as a linear combination of compatible measurements of the complete set.

Let us note that at this level of description, the same as in economics, the subject's physical (or proprietary) capabilities to make a certain decision is an integral part of its state. In the framework of our theory of fundamental measurements, the subject's reluctance or incapability of performing a certain action is not important. The only result of the measurement is its final decision, which actually governs its behavior. The result of application of the theory of generalized selective measurements is the prediction of the statistical regularities of the decisions made by the subject based on the previous measurements. The processes occurring in his or her organism or mind are beyond the scope of the present theory.

Therefore, the "quantum" effects in human consciousness and behavior should be considered only in this meaning. We would like to stress that the "quantum nature of consciousness" in our context relates only to the semantics of its representation in the form of a sequence of elementary selective measurements. At the same time, we are considering only the completed processes of choice (transaction) and action (technology), without analyzing their possible mechanisms. In other words, we limit the set of measured parameters of state only to the obvious macroscopic events. The problems of quantum properties of consciousness analyzed by other authors [6,14] relate rather to the perception of physical mechanisms of choice. The logic of our analysis does not change depending on whether they are quantum macroscopic events or classical events, as we are considering as the elements of consciousness the images and choices already formed by certain mechanisms, i.e. the classical results of brain activity. The necessity of applying the quantum-mechanical formalism only arises in case of an attempt to inscribe these results into the model of a certain enclosed system.

At the same time, the successfully developing theory of macroscopic quantum games [7,8] is the instrument capable of vividly representing the quantum properties of consciousness in the framework of the theory of selective measurements. So far, we have been considering the elementary measurement as an interaction of a classical macroscopic screen (e.g. a semi-screen) with a quantum elementary particle. The function of such a screen can be performed by an employer, whose state practically does not change regardless of the consent or refusal of the transaction of one or even of a certain number of employees. In this case, we can only consider that this macroscopic object is measuring the state of the particle in classical sense. However, if an interaction of two proprietors occurs, and each of them can be considered as an elementary economic object, the interaction between them can be considered as a bargaining, in which their roles are equivalent. At the same time, the result of such measurement (whether the transaction has taken place or not) characterizes the state of the pair of proprietors instead of the states of each of them separately. In terms of the modern theory of quantum information, such states are considered entangled. And only after an additional economic measurement of one of the participants of the transaction we can specify the state of the other participant.

Thus, for instance, if two subjects performed an exchange of a car for a house, we can only state on the basis of this fact that these two objects were of equal consumer value for them. However, in order to find out the consumer value for external observers we should perform an additional measurement of one of them. For instance, we should offer him a certain price for his house. Depending on his consent or refusal of such a transaction, we can specify the information on the "entangled" state of the other subject as well. The theory of quantum macroscopic games primarily deals with this type of entangled states. Let us note that the properties of compatibility of measurements for various subjects can be different as well. This does not result in contradictions in case of considering such subjects as "different particles".

Concerning the actual actions, they act as the technologies that change the consumer properties of elementary particles. By performing a certain action, the subject thus changes the circumstances and receives a possibility of participating in other transactions (or changes the probability of receiving a positive answer in certain elementary measurements). As a result, it turns out that the whole "line of behavior" of the subject can be described as a sequence of measurements and technologies (choices and actions). The task of dynamics in this case is obtaining of the equation (stochastic in the general case), which allows calculating the dynamics of measurement of the subject's state at the specified initial state and external influence. An example of such equation has been obtained by us in the process of analysis of the sequence of choices of a set of traders at an exchange, considered as a result of continuous fuzzy measurement of their quantum state [10]. The obtained system of equations is the quantum-mechanical generalization of the Black-Scholes formula and can be used for decreasing of economic risk.

## 3 Dynamics of economic systems based on the algebra of selective measurements

The dynamics of nonphysical systems is usually constructed phenomenologically by means of formal use of second-order equations and empirical or theoretical determination of effective forces in the system. However, the aforesaid algebra of economic measurements and the related space of states of economic systems allows using a more consistent approach. In the framework of this approach developed by Schwinger changing of the system's state with time is described as a generalized selective measurement. Normal properties of symmetry and continuity of the time axis allow writing down the function of transformation between these two system states in the form of a differential equation of second order (the Schrödinger equation). In this case, formulation of the dynamic law in the form of the variational principle becomes possible as well.

This approach is mathematically common for all systems described in the framework of the theory of generalized selective measurements. Therefore we are going to omit the detailed calculations, making a reference to paper [4], but represent a more detailed discussion of the economic interpretation of notions used in the physical theory. Then we will illustrate the possibility of application of the obtained equations by several simple examples.

### 3.1 Creation and annihilation operators in economic models

In the terminology of Schwinger's algebra of selective measurements the generalized selective measurement $M(a',b')$ can be considered as a compound measurement

$$M(a',b') = M(a',0)M(0,b')) \qquad (10)$$

Here we use the notion of nonphysical empty state «0». At the same time, the first cofactor in (10) can be considered as an annihilation of the system in state a', and the second cofactor – creation in state b'. Special symbols are introduced for these operators:

$$M(0,b') = \Phi(b'); \quad M(a',0) = \Psi(a') \qquad (11)$$

At the same time, the properties of the intermediate zero state are determined by the following formulas:

$$\Psi(a')M(0) = \Psi(a'); \quad M(0)\Phi(b') = \Phi(b') \qquad (12)$$

Measurement $M(0,b')$ means that none of the objects passes through the "input" of this generalized measurement. At the same time, objects in state $b'$ appear at the output. It is difficult to imagine such event in physics, however in economic models it only means that resources, which are not the subject of transaction, are incoming at the input of the technological process. These may include, for instance, air, time, sunlight. At the same time we can state that the value of the manufactured product is created from the zero state. The same refers to the operator of annihilation. It corresponds to the technology of utilization of economic values. Thus, in accordance with the formula (10), any of the indivisible technologies (in which the exchange of intermediate products is impossible) can be represented as a sequence of "annihilation" of resources and subsequent "creation" of products. From the economic point of view this generalized measurement possesses the same properties as the initial one.

From the mathematical point of view such representation allows proceeding from consideration of the algebra of generalized economic measurements to the geometry of state space generated by these measurements.

In the theory of consciousness the operators of creation and annihilation can be interpreted as an origination of a certain attitude to the environment and its annihilation. Actually, the equality (10) in this case means that the sequence of these two events is equivalent to the transformation of such attitude in case if the intermediate states are excluded.

### 3.2 Phase shift as a generalized selective measurement. Continuous description of evolution of state in economic models

The aforesaid notion of the complete set of variables forming a basis relates to static states of the observed system. In Schwinger's methodology the same set of variables in different time points can form different bases. At the same time due to homogeneity of the time scale the transformation function $W_{\delta t} = \langle a';t'|a'';t''\rangle$ between the bases $(a';t')$ and $(a'';t'')$ divided by the interval $\delta t$ generates the transformation function $(W_{\delta t}{}^n)$ for the intervals $n \cdot \delta t$.

Let us note that such transformation corresponds to one generalized selective measurement with input at time point t", and the corresponding equation of dynamics allows calculating the values of coefficients of the transformation function $\langle a';t'|a'';t''\rangle$ depending on the value of t". At the same time the measurement of $a''$ (offer of a certain transaction) does not occur. The only value which is continuously changing is the proprietor's state in the form of possibility of accepting or refusing the offer of transaction $a''$ in time point $t''$. By using the corresponding transformation function $\langle b''|a''\rangle$ we can proceed to the description of state in basis $b''$ and calculate the probability of receiving a consent or refusal of any other offers in time point t".

The decision on which transaction to offer to the proprietor and whether to offer it at the current time point or not, is adopted by the external (in relation to the observed system) observer. The description of this choice requires including it into the system model and describing the extended state from the position of a different "meta observer". Similar aspects occur in the process of discussing the theory of measurements in physics as well, in particular, in the process of discussing the "Schrödinger's cat" paradox.

The assumption of the continuity of the time scale allows writing down the law of dynamics for the infinitely small time increment in the following form

$$\langle a''; t+dt|0; t\rangle = \langle a''; t+dt|a'; t\rangle\langle a'; t|0; t\rangle \quad (13)$$

With account of notation of operators of creation and annihilation of state we can obtain from (13)

$$\Psi(a'', t+dt) = W_{dt}\Psi(a', t), \quad (14)$$

Formal notation in (14) in the form of the differential equation results in the Schrödinger equation

$$i\hbar \frac{\partial}{\partial t}\Psi(a, t) = \widehat{H}\Psi(a, t), \quad (15)$$

Where $\widehat{H} = i\hbar \frac{W_{dt}-I}{dt}$, and $I$ is the unit operator.

As it follows from the principle of scale invariance for the time scale, the obtained operator (Hamiltonian) $\widehat{H}$ no longer depends on the choice of time interval $dt$ and is determined by the properties of the observed system.

Let us note that the rigorous deduction of the differential equation on the basis of (13,14) requires considering the time shift as a variational problem and taking account of all the properties of operator $W_{\delta t}$ and state $\Psi(a', t)$. It has been performed in [4]. In particular, in this process, the connection of the Hamiltonian with the Lagrange operator and the action operator has been derived. Proceeding to the description of state of the observed systems using the density matrix, we can write down the equation of dynamics of an arbitrary set of proprietors using the quantum analog of the Liouville equation (master equation).

$$i\frac{\partial \rho}{\partial t} = \widehat{H}\rho - \rho\widehat{H} \quad (16)$$

The fact of principal importance is that no physical or economic specific features were used for deriving these equations. They are valid for the description of any systems in which the results of observations satisfy the axioms of the algebra of selective measurements. Besides, the general properties of symmetry of the time scale, which are the same for the discussed systems, were taken into account.

The specifics of their use only appear arise when we concretize the type of Hamiltonian. But even in this case significant generalizations are possible. In particular, the properties of homogeneity and isotropy of the coordinate state space allow, in the general case, representing the Hamiltonian in the conventional form as a sum of operators of kinetic and potential energies. Let us note that the aforesaid set of homogeneous generalized selective measurements of value of subject state in the form of coordinate scale ($x$ variable) possesses all the necessary properties of symmetry required for this purpose.

Therefore we will further discuss the examples of economic systems, for which the classical variables of state have a clear meaning. For these systems the Hamiltonian has the following form

$$H = -\frac{\hbar^2}{2m}\nabla^2 + V(x) \quad (17)$$

where x, m and V(x) are the economic analogs of coordinate, mass and potential energy, introduced by us earlier [6].

### 3.3 Variational principle as a basis of constructing the laws of dynamics in physics, economics and in the theory of consciousness

The history of development and application of variational principles in physics is not logically consecutive. The starting point for their discovery was the principle of virtual displacements used for simplified solution of problems of statistics in classical mechanics. The point is that in case of a complex system of kinematically-connected $n$ material points the solution of n force balance equations for each point and n(n-1) equations of paired links between them is required. The total number of unknowns includes n² internal forces (reaction forces). The use of the principle of virtual displacements (requirement of equality to 0 of operation in the process of small virtual displacements allowed by kinematic links) allows reducing both the number of equations and the number of unknowns to the number of degrees of freedom of the discussed system.

In the classical model of a complex economic system discussed by us earlier [12] the equivalent of such kinematic links is the matrix of technologies. Eigenvectors of this matrix correspond to stationary states of isolated system. In case of presence of external influences on the system, using the principle of virtual displacements, we can write down the equations of motion in the generalized coordinates and calculate its trajectory. The next stage of generalization of the principle of virtual displacements is the d'Alembert principle, according to which the problem of dynamics is considered as a problem of statistics in the intrinsic frame of reference by introducing virtual forces of inertia. In this formulation the equations of motion are no more than tautology, as in the intrinsic frame of reference the observed object is at rest by definition. The balance of forces (including the inertia forces) is ensured automatically. The intentional (non-trivial) essence of this principle is rather in the fact that both the external forces and the kinematic links between the material points of the system do not depend on the selection of the reference frame. This invariance allows substantially simplify both the notation of the law of motion and its solutions.

Integrating by time the obtained equation, we obtain the principle of least action. From the mathematical point of view it is equivalent to the requirement of equality to 0 of all forces (including virtual) at each moment of motion. However, it formulation allows deriving the trajectory of motion as a solution of the variational problem. Compared to previous formulations, this formulation is more compact as it requires definition of only one scalar (action) for each of the alternative trajectories of motion. This results in an illusion that the material point independently "selects" the optimum path, i.e. as if it is "looking into the future". However, in reality the principle of least action is valid only locally in the proximity to the trajectory in question and is equivalent to the requirement of instantaneous balance of forces.

The situation in quantum mechanics profoundly changes the attitude to this principle. On the one hand, it can be derived in the same way as in classical mechanics ([4], for instance). On the other hand, the possibilities of delayed choices are taken into account in the current state of proprietor. The reason why all the possible alternatives of the delayed choice influence his strategy is that he has a documented right to choose any of them.

Paraphrasing the classical d'Alembert principle, we can state that the evolution of wave function of an electron passing through the screen with two slits is determined by the balance of forces of action of each of the slits and by the inertia force. The paradoxicality of the two-slit experiment is that in the process of measurement of the coordinate of the passing electron we can find out through which of the slit it has passed. Then it becomes unclear how could the second slit influence its state.

In the economic analog of the two-slit experiment this paradoxicality no longer exists, as it is obvious that the proprietor's strategy will be determined by both alternative possibilities of choice until he is required to reveal his choice. The collapse of state of the proprietor occurs not after "passing through the slit", but at the moment of answering the question – through which slit he has "passed"? It is interesting to note that a similar effect is discovered in the physical two-slit experiment as well (Wheeler's experiment with delayed choice).

Thus, we can consider the principle of least action in economic models as the main variational principle equivalent to the condition of balance of economic forces (including the inertia force) at each moment of time. At the same time, in case of transaction with delayed choice, it is necessary to take account of all the possible alternatives (which are not yet formally rejected by the proprietor) for the calculation of action.

Let us also note that the proprietor is not making any choice in the process of the evolution of state described by the equation (17) or the principle of least action. He simply remains in the framework of the optimum strategy of behavior (balance of forces) at each moment of time. He makes a real choice only when concluding a transaction or refusing of it. This event is describes in the quantum-mechanical formalism as a projective postulate and corresponds to the instantaneous collapse of the wave function of state.

In real systems the refusal of idealization of instantaneous choice results in the necessity of transition to the generalized form of quantum-mechanical formalism, taking account of the possibility of performing continuous fuzzy quantum measurements. Such situation, in our opinion, is more actual in the discussion of the theory of consciousness when the subject does not make a final choice, but decides in favor of a certain alternative of behavior.

### 3.4. Approaches to understanding the meaning of the economic analog of the Planck constant

We have shown in the previous chapters that the results of observation of dynamics of economic systems can be represented as a sequence of generalized selective measurements in the form of transactions and technologies. The probability of obtaining specific results of transactions is determined by the projective postulate and the evolution of state between the transactions is determined by Schrödinger equation. In order to use it, besides the classical variables, the knowledge of the coefficient, which includes the economic analog of the Planck constant, is required. In this chapter we will discuss its meaning and possible methods of its determining.

Let us note that in physics the Planck constant is a measurable value. The principle point for understanding is not its specific value, but the fact that in all experiments it has the same value. In physics this fact is the fundamental law of nature. However, in economic models and in the theory of consciousness we connect the quantum effects with the subjective state of each of the observed proprietors. So can we expect that in this case the experiments on measurement of the economic constant will give the same results as well? One of such experiments is the studying of diffraction properties of a set of proprietors in case of "passing" them through an economic slit of variable width [3]. In the process of its gradual expansion the number of passing economic particles increases, first proportionally to the square of width of the slit and then linearly. The exact measurement of this dependency allows calculating the effective wavelength of particles, and consequently, the Planck constant (classical values of the economic mass of particles are calculated or measured independently).

Another method of measurement of the Planck constant is the observation of the energy spectrum of the system located in a potential well. Below we will discuss this model in detail. In the process of such measurement the energy spectrum lines appear to be diffused by order of magnitude $kT$. Therefore, the values of the Planck constant will match only with the accuracy of measurement of these levels. It is quite expectable that in economic models the level of thermal noise makes a significant contribution to the results of measurement of state. We assume that all not accounted properties of the subject (its affections, skills, limitations, delusions, etc.) can be considered as a thermal noise, which is not taken into account in the analysis of the results of the main experiment. At the same time, as we approach the ideal state and zero value of the thermal noise, we will be obtaining more and more indistinguishable in economic sense states of consciousness, and more and more accurate value of the measured Planck constant.

Another important factor, in our opinion, is the necessity of comparison of dimensionless quantities or values having the same dimension. Formally, such identity of units of measurement can be ensured only by their additional comparison. In other words, measurements of various economic experiments (for instance, the quantity of bananas offered for exchange) must be directly or indirectly interconnected by an additional economic measurement (transaction of exchange of different batches of bananas). At the same time, each of the proprietors, whose state is being measured, will have a possibility to use this additional transaction for receiving riskless profit (arbitrage). Idealization, in which the possibility of arbitrage is absent, is apparently the factor, which equalizes the values of variables obtained in the process of observation of various economic systems.

The aforesaid arguments are rather "reasonable arguments" than an explanation of the economic essence of the economic constant. But our partial excuse is the fact that in physics this problem also remains unsolved.

Next, we will discuss the models of simplest economic systems and describe some of the features of their dynamics on the basis of the proposed approach.

## 4 Dynamics of simplest models in the theory of consciousness and in economics

At the initial stage of construction of the theory we are not setting a task of discovering in economics or in the theory of consciousness any structures, which have no analogs in physical models. We are going to use the already known solutions of the Schrödinger equation for the specific type of Hamiltonian. Nevertheless, even the application of properties of mathematical models already studied in physics to the economic models requires a thorough analysis. The point is that the majority of physical parameters and variables of state in economic models do not have any obvious analogs and require substantiation in each specific case.

Besides, the aim of our discussion of simplest models of economic systems is to illustrate certain quantum properties, which cannot be explained in classical models of economic systems and appear to be paradoxical.

One of the first observable quantum effects is the phenomenon of interference. Though it had been discovered long before the origination of quantum mechanics, and the wave theory had been used for its substantiation, it eventually became clear that it can be considered as a quantum effect and the light wave - as a photon flux. Such effect for the simplest unidimensional motion of particles is the nonlinear dependency of the light transmission coefficient on the thickness of semi-transparent film. For obtaining the correct result we need to consider the superposition of an infinite number of alternatives, each of which corresponds to the finite number of light reflection from the film surfaces.

### 4.1. Classical unidimensional motion of economic "material point"

Before describing the economic analog of such experiment, let us discuss the classical model of unidimensional motion through a potential barrier. For a flying photon the thin film is a domain, in which the velocity of light decreases in accordance with the parameters of refraction. In case of unidimensional motion of a classical particle, such local decrease of speed corresponds to a domain with greater potential energy (Fig.5). All classical particles with kinetic energy greater than the height of the potential barrier roll over it, and the rest reflect from it. At the same time, the shape of the potential hill makes no difference.

We have previously [12] determined the following kinematic and dynamic parameters for the classical unidimensional motion of the economic "material point":

- Coordinate $= \log_2 C/C_0$, where $C$ is the complete capital, $C_0$ is the capital corresponding to the origin of the reference scale.
- Pulse $p = s - s_0 = \delta s$, where $s$ is the cost of expenses for manufacturing of a unit of product in accordance with the technology, $s_0$ is the price of sale of a unit of product; $\delta s$ is the surplus value of manufacturing of a unit of product
- Force is the pulse rate
- Mass $m = C/P$, where $P$ is the productivity
- Law of motion $\dfrac{C}{P} \cdot \dfrac{d^2}{dt^2} \log_2(C/C_0) = \dfrac{d}{dt} \delta s$

It follows from this model that the potential function $U(x) = \int_0^x F(x')dx'$ can be introduced in case when $F = \dfrac{d(\delta s)}{dt} = f(x)$. In other words, when the rate of changing of the surplus value per unit of product is determined only by the total capital or the productivity proportional to it in case of simple extended reproduction.

At the same time, all the internal features of the technology remain constant for the material point. Therefore, only one scalar parameter of the internal state of object is included in the equation of dynamics - the mass of object. In this case, the law of surplus value acts as the second Newtonian law (or the definition of force). In case of motion of such system, its complete mechanical energy is conserved.

Assume, for instance, that when the capital of a certain company reaches the value $X_1$, the market of products reaches the balance. Then, in case of further increase of the capital and, correspondingly, increase of productivity, the unit price of the manufactured product starts to decrease. If the rate of this decrease depends only on the productivity (its increase over the equilibrium value), then the condition of potentiality is satisfied, and the dynamics of the system corresponds to the motion of a ball, rolling over a hill in a gravity field (Fig.6).

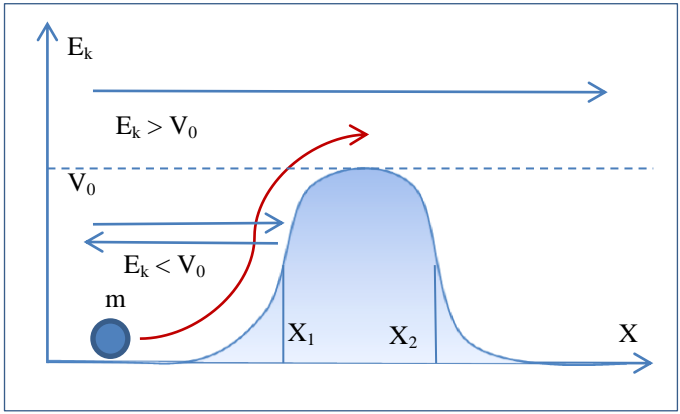

Fig.6 Diagram of classical unidimensional motion of an economic material point through a potential barrier

In this case, the speed of system motion decreases and when the company stops to bring profit it becomes equal to zero. However, the capital and the productivity are at maximum at this moment and the product price continues to decrease. At this moment the company formally begins to work at a loss. Its capital, productivity and coordinate decrease with increasing speed until the equivalent ball rolls down the hill. When it returns to the initial point, the product price becomes lower than the initial price, which corresponds to a negative value of the impulse.

The situation described above does not claim to correspond to the real dynamics of product prices; however, it illustrates the motion, which originates in case of observing the aforesaid condition of potentiality.

The descent following the rise may mean that the price for recourses required for manufacturing begins to change (fall) in the same manner. Like in the first case, it is connected with the increase of volumes of production and, correspondingly, the volume of purchases. If the economical material point manages not to stop on the way to the top of the potential barrier (its kinetic energy calculated according to general formula exceeds its height), then it continues its motion.

**4.2 Interference of states (delayed choice)**

It is assumed in the described classical model that the cost of the company (complete capital) and its profitability have simultaneously objective values. However, we have previously shown the logical inconsistence of such assumption. Therefore, for a more correct description of the dynamics we should consider that both the capital of the company (cost of its sale) and its profitability (estimated speed of cost changing) are the results of the generalized selective measurements – answers to the offers of transactions.

At the same time, it is appropriate to consider a "hill" with vertical walls corresponding to the boundaries of a semi-transparent film in the interferential experiment. In this case, at the moment of crossing of each of the boundaries by the object, we exactly know the values of the coordinate corresponding to the complete capital (the moment of crossing the boundary is not determined in this case). Besides, due to the fact that the "potential hill" is caused by macroscopic external influence, its height can be set as accurately as desired. Using the law of mechanical energy conservation, we can calculate the exact value of the pulse of the object both in case of crossing the boundary, and in case of reflection from it. In both these case the quantum principle of uncertainty is violated, prohibiting the simultaneously known values of coordinate and pulse. In the discussed case we have only two alternatives of motion. Therefore, the principle of uncertainty can be satisfied only by allowing both alternatives with a certain degree of probability.

In the economic interpretation it means that if a proprietor is offered a transaction, in which the "measuring slit" corresponds to the exact value of the cost of his company (boundaries of the barrier), he will "pass through the slit" only in case of availability of two weighted alternatives of the value of profitability. In modern economic relations such possibility is available in the form of various types of options. The simplest of them is the futures contract - a right to buy out a share in the future at its current price. It is obvious that the price of such futures contract is equal to the expected value of profitability of the share.

Thus, a portfolio consisting of a share and a simple futures contract for it represents the simplest example of the generalized selective measurement with delayed choices. Two alternatives of such choice mean that the proprietor may either buy out the share in the agreed period of time, or refuse to buy it. In this case, the share price (coordinate) is measured exactly, while its profitability is undetermined. The price of futures contract depends on the degree of certainty of the proprietor in the fact that he is going to buy out the share upon the agreed period.

Let us note that the discussed portfolio consisting of the share and the futures contract is only used for illustrating the measurement with delayed choice. Exact equivalence to the physical experiment of penetration of the semi-transparent screen by particles can only be provided by a special selection of conditions of transaction; formulation of these conditions requires a more thorough analysis. We are currently illustrating only the possibility and imminence of such situations in economics.

The second boundary of the potential barrier can be of different shape and height. However, in case if it represents specular reflection of the first boundary, we can consider its penetration similarly to the first boundary, but with a reversed sign with respect to time. And if the proprietor is offered to buy a certain option (or a portfolio with option) at the first boundary, at the second boundary he will be offered to sell it.

When two such offers with a delayed choice (option) are available, the situation, in which the alternatives of choice interfere, occurs. For instance, it is possible to refuse buying out a share (sell the option) only in case of preliminary agreement for it (by means of buying the option). As a result of performing twos such sequential selective measurements the proprietor's state is changing. But the choice itself (buying out the share or refusing of it) turns out to be not needed. In this case, the proprietor's state is described not by the mix of two possibilities with various degrees of probability, but by their superposition.

In the equivalent physical experiment (passing of the photon through the semi-transparent film) the possible trajectory of its motion (Fig.7) is unknown and we cannot state that it passed along any of these trajectories. In other words, there are no parameters hidden from the observation corresponding to a particular trajectory of motion. Verification of Bell inequalities excludes such possibility. In the economic model this means that the proprietor's state (and behavior) cannot be described by a mix of strategies of behavior in case of a particular choice. It corresponds to the situation when the delayed choice remains not needed. The result of such selective measurement is the consent for both transactions (purchase and subsequent sale of the option) or the refusal of both transaction (Fig.7). However, in this case, the choice is not required whether to buy out the share according to the conditions envisaged by the option, or not.

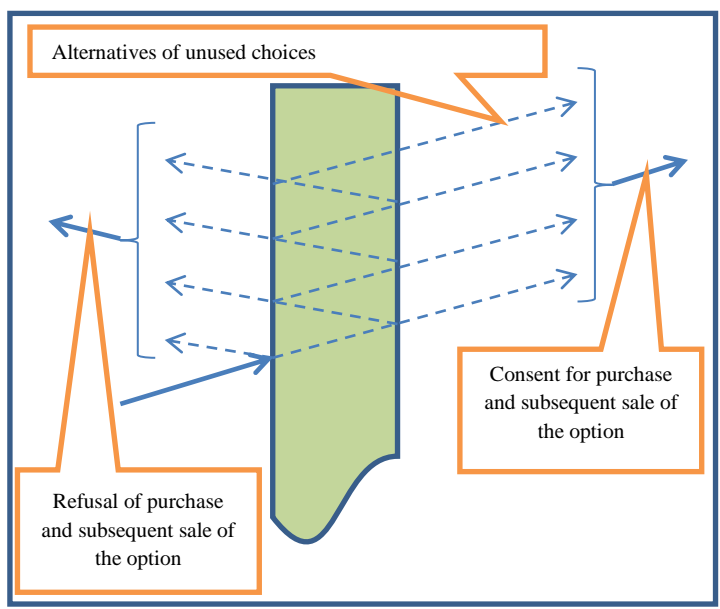

Fig.7 Diagram of the superposition of alternatives in the economic model with two delayed choices may mean the absence of choice. It is equivalent to the model of passing of a microparticle through a semi-transparent wall of finite thickness.

**4.3 Methods of modeling unidimensional dynamics in economic systems**

In the previous chapter we have discussed in details the qualitative equivalence of the interference of states of the delayed choice to some simple physical models. For obtaining the quantitative estimates of parameters of this state we no longer need to calculate the interference of the proprietor's strategies every time, as it is done in the theory of macroscopic quantum games [7].

We can use the Schrödinger equation obtained from general considerations in the theory of generalized selective measurements. It can be considered as a simplified description of the dynamics of state of the proprietor, which is possible due to the symmetries of the state space and time. Ordinary classical formalism of calculation of probabilities of his alternative strategies and choices is also possible, but it requires a more complex description and a significant increase of the number of variables of state [3]

The method of using the quantum-mechanical formalism for the analysis of dynamics of unidimensional single-particle economic models is based on the following stages:

- Introducing a variable of state as a coordinate – logarithmic scale of object value (for a company – logarithm of its total capital)
- Calculating the mass parameter – the capital output ratio of manufacturing of a unit of product (this parameter is defined only by the technological parameters of the manufacturing process and is considered classical and constant in the discussed model)
- Independently obtaining (from economic or statistical models) the dependency of the rate of change of the surplus value per product unit (force) on the capital (coordinate). It is always possible for systems, in which the relation of demand and supply depends not on the price of products and resources, but on the rate of its change. Constructing the potential function $V(x)$.
- Calculating the effective value of the constant (economic analog of the Planck constant) in equation (15) based in the additional test measurements.
- Solving the equation (15) or (16) for the set initial conditions in relation to the wave function $\Psi(a,t)$.
- Calculating the probability of the proprietor's consent or refusal for the offered transaction using the projection postulate.
- Without concretizing the initial conditions, it is also possible to calculate the main generalized parameters of dynamics, such as stable states, transmission coefficients, etc.

Let us note that all the aforesaid parameters relate to the time interval between the two selective measurements (offers of transactions and receiving answers for these offers). In the general case, the procedure of observation of the economic system can be quasicontinuous and should be described by the generalized formalism of continuous fuzzy quantum measurements [16]

**4.4 The expected quantum-mechanical effects in unidimensional single-particle economic models**

In the discussed rather narrow set of models all quantum-mechanical effects have already been well-studied and described in numerous physical applications. All of them are described by the solutions of the Schrödinger equation [15]. Therefore, in this chapter we are only to list going the main of them, indicating the economic specifics and possibilities of their observation.

- *«Interference on thin films»*.

Energy analog of the thin film corresponds to the potential barrier in case when the energy of the particle exceeds its height. Such barrier in economic models appears when a transaction with delayed choice is compensated by another similar transaction. At the same time, the choice stipulated in the conditions of the transactions (buying out of a share, for instance) is not realized and the calculation of the proprietor's state is made with account of all the possible alternatives. The shape of the potential barrier depends on the conditions of transactions and is calculated in accordance with the classical formulas. The width of the barrier depends on the time interval between the first and the second transaction. The main quantum effect is the dependency of the transmission coefficient (probability of consent for the pair of transactions) on the width of the barrier. For sufficiently small values of width it should be of oscillatory nature. The oscillation period corresponds to the effective wavelength of the economic "material point". On the basis of it, the value of the "economic Planck constant" can also be calculated.

- *«Slit experiment»*

Definition of the subjective value of price of property (in the units of certain valuables) presupposes a specially-organized transaction (auction), which allows finding out which of the participating proprietors assume that the price of their property is within a certain range. The probability of passing through such slit increases with the increase of its width. In physics this dependency for sufficiently narrow slits (with width smaller or equivalent to the de Broglie wavelength of the passing particle) is quadratic. In case of subsequent increase of the slit width this dependency becomes linear. In our paper [3] we have analyzed in detail the economic analog of this experiment and have shown that it can be used for determining the economic analog of the de Broglie wavelength of the proprietors passing through the economic slit. Besides, it has been shown that the description of such measurement allows a classical probabilistic, but it requires the use of the category of conditional choice. In this connection, the possibility of quantum-mechanical description can be considered as a simplified description possible due to the symmetries of the coordinate and time axes.

- *«Quantum harmonic oscillator»*

One of the most noticeable quantum effects in physics is the quantization of energy levels in the harmonic oscillator. We have previously described in detail an example of occurrence of the classical harmonic oscillator in the island model [12]. It illustrates the interaction of two tribes of "hunters" and "farmers". The main dynamic variable is the distance between the coordinates of the tribes – the logarithm of ratio of their conditional capitals.

$$x_2 - x_1 = \ln(NC_2) - \ln(NC_1) = \ln\left(\frac{NC_2}{P_2}\right) - \ln\left(\frac{NC_1}{P_1}\right) - \ln\left(\frac{P_1}{P_2}\right)$$
$$= m_2 - m_1 - \ln\left(\frac{P_1}{P_2}\right) = (x_2 - x_1)_0 + \ln\left(\frac{P_1}{P_2}\right)_0 - \ln\left(\frac{P_1}{P_2}\right) \quad (18)$$

In case when the masses of the corresponding material points are fixed and known it is connected with the ratio of their productivities. It is obvious that this variable of state is classical and its exact value can be measured, the same as the speed of its changing.

Nevertheless, from the economic point of view, it is only a technical parameter and it does not define the cost of the "company". Even if we know the exact amount of manufactured products, its surplus, and the proportions of exchange between the tribes, there are still several alternatives of transactions, which may result in such exchange. In the classical model there are no discrepancies about the calculation of the distance between the points as we define the price as a proportion of exchange.

However, a "conscious" tribal chief will be able to calculate the subsequent oscillations of the productivity. In order to secure himself for the future, he agrees with the other chief that he will deliver the products insufficient for the "fair" price later, when there will be a demand for them and they will be produced. In this case, the price is determined not only by the proportion of exchange, but also by the obligations undertaken by each of the tribal chiefs.

However, in this case, the price remains measurable and determined by the conditions of the transaction. As we have previously ascertained, the quantum properties occur only in case when no choice between the alternatives is made and its possibility is neutralized by another delayed choice. In the discussed example such delayed choice is any right that is not an obligation, i.e. an option.

From the formal point of view it is appropriate to consider a potential well with vertical walls first, as an inversion of the aforesaid potential barrier. Alternatives of the trajectories of the particle in such well correspond to Fig.7, but the kinetic energy of the particle allows it to be located only within the set interval, not beyond it. This means that none of the alternative trajectories not within the interval is is admissible. If we consider a bounce of the particle from each of the walls of the well as a selective measurement with a delayed choice, then the sum of several such measurements may create a situation when the choice will not be required, as the alternative possibilities of choice will compensate each other. Using the terminology of wave mechanics, we can say that the alternative selective measurements (offers of transaction in economic context) compensating each other are in the antiphase.

Unlike the semi-transparent screen, only particles, for which the possibility of being detected beyond the boundaries of the infinitely high walls is completely impossible, can be located in the potential well. At the same time, the alternatives in the state of antiphase can follow one another, or alternate. In physical models a certain energy level of the well corresponds to each of such possibilities. For the harmonic oscillator the shape of the well is set in the form of a parabola, ensuring constant distant between the energy of stationary states of the system.

In an economic potential well the proprietor's state is described in a similar way. While the proprietor is in a certain stationary state, he is ready for an exchange of property at any price from the limited interval with a certain probability. At the same time, the density of probability distribution is not changed with time. For maintaining this state the proprietor must conclude a number of transactions. The rights and obligations (delayed choices) obtained by him as a result of these transactions <u>in aggregate</u> guarantee the impossibility of changing of the price of property beyond the limits of the mentioned interval. Representation of this summary transaction in the form of an infinite sum of alternative trajectories (Fig.7) is only a method of series expansion of the proprietor's stationary state into various classical non-stationary states.

We will not lay out the specifics of the terms and conditions of these transactions, as a great number of equivalent methods of their representation exist, and the results of possible observations can eventually be predicted on the basis of the solutions of the equation (15) with a potential corresponding to the shape of the well. Let us only note that the oscillatory nature of the dynamics of technological parameters of production allows determining the classical parameters in the equation (15), but it is not directly connected with the wave nature of the description of the proprietor's stationary state.

For observing the predicted quantum effects it is sufficient to calculate the analog of mechanical energy of economic objects located in the potential well in the stationary state. We can also assume that the influence of external random factors is equivalent to temperature noises in the observed system. At significant noise level we can expect the smearing of levels up to complete disappearance of quantum properties of the spectrum.

The observation of the effect of resonance absorption in such systems is also possible. It occurs as an induced transition of the system from one state to another in case of external influence of a particular frequency.

- *«Tunnel effect»*

The tunnel effect in physics occurs when the kinetic energy of a particle is insufficient to overcoming the potential barrier in the classical way. Otherwise, this situation is similar to the aforesaid situation of passing through the semi-transparent screen.

In the economic model it means that there is no classical state equivalent to various alternatives of the aforesaid pair of measurements-transactions with a delayed choice. Similarly to physical models, we can measure the coordinate of the particle at the moment of tunneling and with a certain probability obtain the value corresponding to the under-barrier domain. However, in this case, the formal calculation of the kinetic energy gives a negative value, which means that the pulse of the economic particle is imaginary, unlike the over-barrier passing, when we can calculate the value of the pulse, and its uncertainty is connected with the superposition of direct and reflected alternatives (Fig.7). Not claiming economic correctness, we assume a connection between such imaginary pulse value (profitability per unit of product) and, accordingly, the speed of motion, and the impossibility of measuring their value on the discussed segment of the coordinate scale. We can also assume that in this case the proprietor's state does not allow him to participate in transactions of purchase and sale of option for his property.

As before, studying the dependency of the transparency coefficient for the under-barrier tunneling on the height and shape of the barrier, we can also determine the effective values of all parameters of state of the set of the observed economic particles and the value of the economic analog of the Planck constant.

- *Continuous fuzzy measurements in stock exchange dynamics*

In our paper [ ] we have obtained the quantum-mechanical generalization of the Black-Scholes formula for calculating the dynamics of the price of options. At the same time, we have used the generalization of the quantum-mechanical formalism for the model of continuous fuzzy quantum measurements. Its specific feature is the refusal of the idealization of instant acquisition of information on the results of measurements. For multiparticle weakly coupled quantum systems such model allows writing down the equation of dynamics in the operator form without detailing of the mechanics of interaction, only based on the information characteristics. Here we are only going to illustrate the system of equations describing the quantum-mechanical analog of the Black-Scholes formula (19) and its classical analog (20). Unlike the previously discussed artificial simplified models, the model of stock exchange dynamics can be directly used for the analysis of real economic systems.

$$\left[ \begin{array}{l} \dfrac{\langle \hat{f}\hat{C} \rangle}{\langle \hat{f}\hat{S} \rangle - \langle \hat{f} \rangle \langle \hat{S} \rangle} = \dfrac{\langle \hat{S}\hat{C} \rangle}{\langle \hat{S}^2 \rangle - \langle \hat{S} \rangle^2} \\ \left\langle \dfrac{\partial \hat{f}}{\partial t} \right\rangle + r\langle \hat{S} \rangle \dfrac{\langle \hat{f}\hat{C} \rangle}{\langle \hat{S}\hat{C} \rangle} + \\ \dfrac{\langle \hat{f}\hat{B} \rangle \langle \hat{S}\hat{C} \rangle - \langle \hat{S}\hat{B} \rangle \langle \hat{f}\hat{C} \rangle}{\langle \hat{S}\hat{C} \rangle} = r\langle \hat{f} \rangle \end{array} \right. \qquad (19)$$

$$\dfrac{\partial f}{\partial t} + \dfrac{1}{2}\sigma_S^2 S^2 \dfrac{\partial^2 f}{\partial S^2} + rS\dfrac{\partial f}{\partial S} = rf \qquad (20)$$

**4.5 Behavior as a continuous fuzzy measurement**

The aforesaid economic models can be also used for modeling of subject behavior. We have already noted the analogy of the dynamics of economic "material points" and the human consciousness. All the aforesaid expected quantum effects in economic systems can be observed in human consciousness as well. In particular, a lot of examples of discreteness of states of consciousness can be found. By learning to calculate the energy of these states and measure the effective value of the analog of the Planck constant, we will be able to facilitate the transition of consciousness from one state to another using the resonance method.

Any of the subject's choices can be considered as an answer for an offer of transaction, in which certain values are at stake (both positive and negative – "carrot" and "stick"). The only difference is that this kind of "transactions" can be offered to a human not only by other humans, but also by objective circumstances of his destiny. The state of consciousness changes not only in case of consent for a transaction, but also in case of refusal. The choice is followed by a certain action, which can be considered as a technology-type generalized selective measurement. Quantum effects of consciousness are revealed in situations of "delayed choice", when the adopted decision allows alternatives of behavior. The final choice of alternatives can remain unrealized if the subsequent choices interfere with it.

Such scheme of dynamics of consciousness is simplified. In real life the human consciousness simultaneously makes several of choices at any moment of time. At the same time, most of them can be characterized as delayed choices. Besides, the choice itself can be prolonged in time. At the same time, the subject's degree of certainty of the fact that it is necessary to accept one of the alternatives changes gradually.

We account of the aforesaid, we can conclude that the most general model of conscious and behavior of subject must be multiparticle, multidimensional and must satisfy the conditions continuous fuzzy measurement.

**Conclusion**

The proposed approach of modeling the dynamics of economic systems and human behavior has principal differences from the generally accepted concepts. It is based on general scientific conceptions assuming that the state of the observed object is completely and unambiguously determined by the results of measurements performed on it. In this case, the general form of notation of the laws of dynamics is determined by its specifics, and the properties of the object itself only determine the parameters and the specific form of functions included in the equations.

We consider the offer of transaction of exchange of valuable items as a fundamental measurement for economic systems including the consciousness of a proprietor as an integral part. Thus, we exclude from the analysis the discussion of various mechanisms of consciousness and describe its properties on the basis of the received answers. A proprietor's consent or refusal of a transaction is the result of measurement of his state. In the framework of this approach we have constructed the algebra of fundamental economic measurements. It has been shown that any indivisible transaction of any complexity can be considered as a formula of Boolean algebra and represented in the form of a disjunctive normal form of fundamental economic measurements. At the same time, not only any measurement represents an influence on the observed system (changes the proprietor' state), but also, vice versa, any influence can be considered as a certain generalized measurement of parameters of his state.

It has also been shown that the Schwinger's algebra of selective measurements, on which the derivation of the quantum-mechanical formalism was based, can be derived as a particular case of the algebra of fundamental economic measurements. For this purpose, it is necessary to ensure conditions of a specific type of transactions, allowing the introduction of the scale for homogeneous measurements. A demonstrative screen interpretation for such transactions has been proposed.

We do not use any specific properties of specific economic systems for the derivation of the equations of dynamics of economic systems. The represented examples are only the simple idealized illustrations of the proposed approach. However, the properties and structure of the performed measurements are of principal importance.

Their general analysis allows showing that the economic systems can (and in many cases must) possess quantum properties. The point is that the availability of consciousness as a component of the idealized part of the system allows it to "predict" the determined future and make corrections to the answer for the offer in accordance with this prediction. As a result, only self-consistent answers for the offers of transactions become possible (results of fundamental economic measurements). Besides, the exact values of company price and its profitability (economic analogs of coordinate and speed) can be self-consistent only in the degenerate case of zero profitability and continuous observation, which corresponds to the Zeno quantum effect in physics. The rest of the cases result in a state, for which these parameters cannot be simultaneously precisely determined (analog of the principle of uncertainty)

It has been shown by us that in case when a proprietor agrees or refuses of a transaction with a delayed choice, the properties of the systems are quantum-mechanical. For instance, when a proprietor is purchasing an option, he pays not for the security itself, but for the right of buying it out at the agreed price later. Until the choice between the alternatives (to purchase the security or to refuse to buy it out) is made, the proprietor's state is determined not by their weighted mix, but by their superposition. The dynamics of such systems is described by the Schrödinger's equation, and the calculation of the probability of consent or refusal of any other offer of transaction required the use of the quantum-mechanical formalism.

If two measurements with a delayed choice compensate each other completely or partially, than the delayed choice remains not used. In this case the dynamics of the observed economic system is fundamentally quantum, and cannot be reduced to any of the classical alternatives. For such situations the economic analog of Bell inequalities can be obtained.

Further application of the proposed theory no longer requires the detailed analysis of the mechanism of decision-making of the proprietor. The obtained equations formalize the account of all his features. It is only necessary to determine the values of the classical parameters included in the equations of dynamics on the basis of preliminary observations. Besides, we can expect that the qualitative analysis of the possible solutions will allow consciously constructing new financial instruments providing the occurrence of particular properties.

The final aim of application of the proposed formalism is the maximally complete (within the limits of the principle of uncertainty) control of the dynamics of the observed economic systems.

We would also like to emphasize that the proposed formalism of measurements is applicable not only to economic systems, but also to all other systems, in which the situation of delayed choice occurs. In particular, human behavior can also be represented as a sequence of choices (bargains with circumstances) and technologies (actions realizing the choices). In the general case this sequence can be considered as a continuous fuzzy quantum measurement.